\documentclass[preprint, superscriptaddress, aps, prb]{revtex4}

\usepackage[colorlinks=true, citecolor=blue, urlcolor=blue, linkcolor=blue]{hyperref}
\usepackage{amsmath,amssymb,gensymb}
\usepackage{graphicx}
\usepackage{siunitx,physics}

\DeclareMathOperator{\spn}{span}

\begin{document}
\title{A large-scale first-principles quantum transport simulation method using plane waves}
\author{Meng Ye}
\affiliation{State Key Laboratory of Superlattices and Microstructures, Institute of Semiconductors, Chinese Academy of Sciences, Beijing 100083, China}
\affiliation{Materials Sciences Division, Lawrence Berkeley National Laboratory, Berkeley, California 94720, USA}
\author{Xiangwei Jiang}
\email{xwjiang@semi.ac.cn}
\author{Shu-Shen Li}
\affiliation{State Key Laboratory of Superlattices and Microstructures, Institute of Semiconductors, Chinese Academy of Sciences, Beijing 100083, China}
\author{Lin-Wang Wang}
\email{lwwang@lbl.gov}
\affiliation{Materials Sciences Division, Lawrence Berkeley National Laboratory, Berkeley, California 94720, USA}

\begin{abstract}
As the characteristic lengths of advanced electronic devices are approaching the atomic scale, \textit{ab initio} simulation method, with fully consideration of quantum mechanical effects, becomes essential to study the quantum transport phenomenon in them. However, current widely used non-equilibrium Green’s function (NEGF) approach is based on atomic basis set, which usually can only study small system with less than 1000 atoms in practice. Here we present a large-scale quantum transport simulation method using plane waves basis, based on the previously developed plane wave approach (\textit{Phys. Rev. B} \textbf{72}, 045417). By applying several high-efficiency parallel algorithms, such as linear-scale ground-state density function theory (DFT) algorithm, folded spectrum method, and filtering technique, we demonstrate that our new method can simulate the system with several thousands of atoms. We also use this method to study several nanowires with about 4000 copper atoms, and show how the shape and point defect affect the transport properties of them. Such quantum simulation method will be useful to investigate and design nanoscale devices, especially the on-die interconnects.
\end{abstract}

\date{\today}
\maketitle

\section{INTRODUCTION}
The characteristic scale of advanced electronic devices have reached nanoscale. As the International Roadmap for Devices and Systems (IRDS)\cite{IRDS} shows, the active channel length of an electronic device has been downscale to tens of nanometers. The traditional technology computer aided design (TCAD) method is based on continuum level models, which cannot be applied to such small systems where the atomistic quantum effects are dominate. Instead, atomistic modeling and simulation method based on quantum mechanics become necessary.

In the transport simulation, a two-probe model, which contains two semi-infinity electrodes and a central region for scattering, is widely used. The current in this model can be calculated by using the Landauer-Büttiker formalism
\begin{equation}
I=\frac{2e}{h}\int [f(E-\mu_\text{R})-f(E-\mu_\text{L})]\sum_nT_n(E)\dd{E}
\end{equation}
Here $f$ is the Fermi–Dirac distribution function, $\mu_\text{L}$ and  $\mu_\text{R}$ are the Fermi level of left and right electrodes, and $T_n(E)$ is the transmission coefficient for the $n$-th right electrode band at energy $E$. Since $\mu_\text{L/R}$ can be obtained by using \textit{ab initio} bulk calculation, the key problem of transport simulation is to calculate $T_n(E)$. Currently, method based on non-equilibrium Green’s function (NEGF) theory\cite{Brandbyge2002} is the most widely used approach. Tight-binding model based NEGF approach can simulate a system containing more than 100000 atoms,\cite{Klimeck2007,Klimeck2007a} but it ignores the charge density self-consistent effect which is important at atomistic level. Furthermore, NEGF approach combining with \textit{ab initio} density functional theory (DFT) prefer a localized atomic basis set rather than plane waves. However, a localized basis is less accurate and less flexible than a plan wave one in DFT calculations.\cite{Davidson1986} Although there are several approaches to generate localized Wannier functions from the plane wave calculation, many approximations must be made and many of these method are computationally expensive.\cite{Calzolari2004} The widely used DFT-NEGF approach usually can only simulate system less than 1000 atoms, which is not enough for a real nanoscale device.

Given these circumstances, it will be interesting to test different methods. One approach is to directly solve the scattering states using plane wave basis and auxiliary periodic boundary conditions.\cite{PhysRevB.72.045417,Garcia-Lekue2006,GARCIALEKUE2015292} This approach can be easily applied to the existing DFT code, and the computational resource it takes is similar to that of a standard ground-state calculation.\cite{PhysRevB.72.045417} However, since the computational complexity of common DFT method is $O(N^3)$ (here $N$ is the scale of system), and many eigenstates of the system need to be calculated, it can only be used to study the system contains hundreds of atoms in practice. In this work, we combine this plane waves simulation method with linear-scale ground-state DFT algorithm, as well as the folded spectrum method\cite{doi:10.1063/1.466486} and filtering techniques, to solve problems with thousands of atoms. In particular, we study on-die interconnect of copper nanowires, with several thousands atoms, and show the shape and point defect effect in them. Our new method will be useful to study large-scale nano-electronic devices, especially for nanoscale interconnects.

\section{COMPUTATIONAL PROCEDURE}
\begin{figure}[htbp]
\centering
\includegraphics[width=8.5cm]{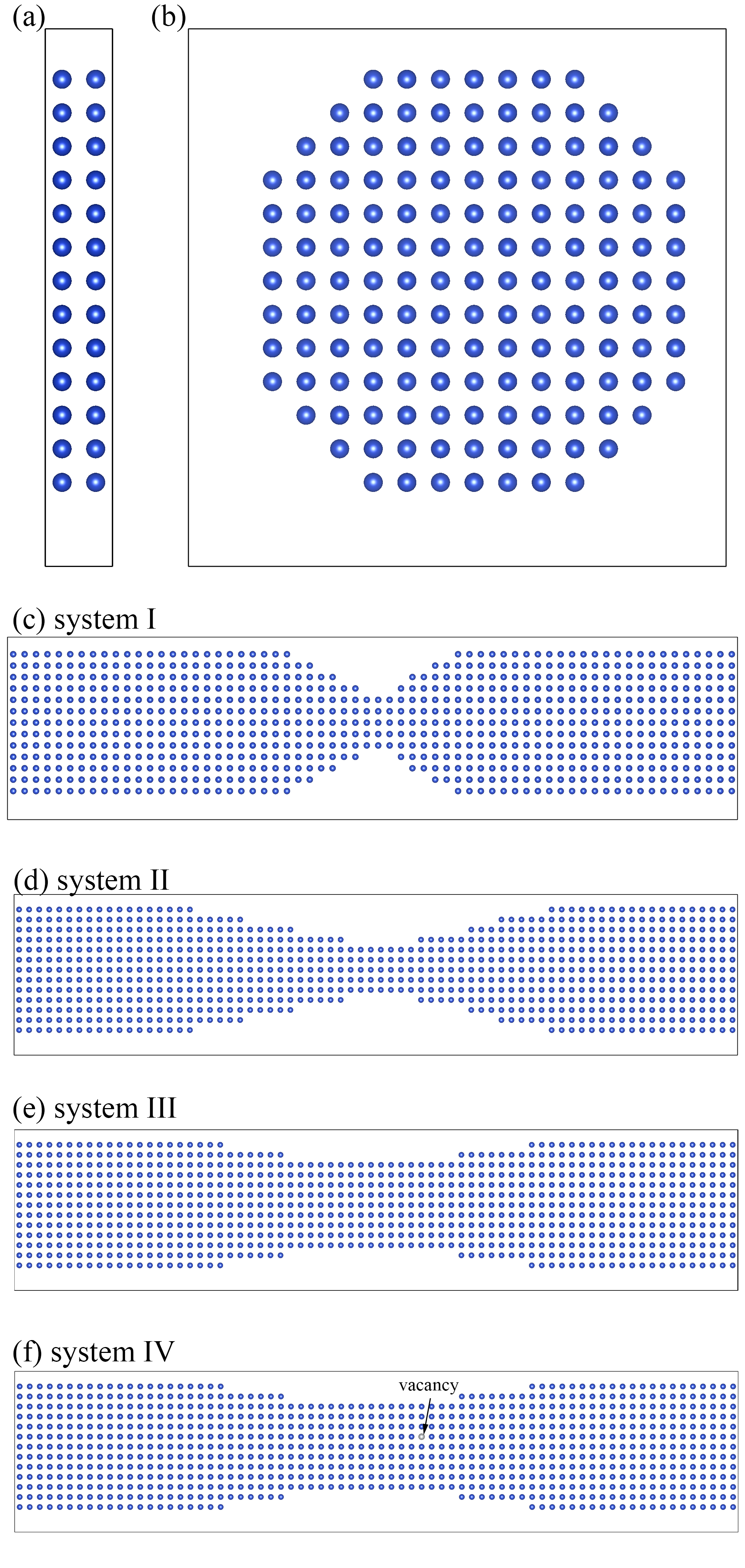}
\caption{(a) and (b) are the side and top view of electrode of all system. (c)-(f) are the side view of system I-IV, respectively. The arrow in (f) indicate where the vacancy defect located.}\label{fig:struct}
\end{figure}
 
The system we study in this work is copper nanowires, which are commonly used for on-chip interconnects of nowadays integrated circuit (IC). And such interconnect is the main source of heat generation, is also facing difficulty for further downscaling. In this work we use four systems with the same electrodes(which is a copper nanowire with radius of \SI{1}{nm}, and its cross sections are shown in FIG. \ref{fig:struct}(a), (b)) to study. The structure of four systems (which are called system I, II, III, and IV) are shown in  FIG. \ref{fig:struct}(c)-(f), the transport direction are all along the $[100]$ direction of bulk copper, which is the $x$-direction in these structures. The main difference between system I, II, and III is radius and length of narrow part (we call it ``neck'') in the central region, while system IV is created by introducing a vacancy defect in system III (as the arrow in FIG. \ref{fig:struct}(f) indicate).  The number of atoms in these four systems are 3996, 3660, 4140, and 4139, respectively. The good to use these system is to investigate how the  central part shapes and point defect affect the interconnect transmission.

In all the calcultions of this work, the local density approximation (LDA) with the Perdew-Zunger parameterization\cite{Perdew1981} of the exchange-correlation functional and Fritz-Haber-Institute (FHI) norm-conserving pseudopotentials\cite{Fuchs1999} are used, and the cutoff energy is set to \SI{40}{Ryd} ($\sim\SI{540}{eV}$). All the calculations are done on the Titan supercomputer\cite{Tian}, while the central processing unit (CPU) and graphics processing unit (GPU) are AMD Opteron 6274 and Nvidia K20X, respectively.

\subsection{The original approach}
The central idea of the plane wave approach is to get scattering state wave funtions $\psi_\text{sc}$ by solving the Schrödinger equation $\hat{H}\psi_\text{sc}=E\psi_\text{sc}$ ($E$ is the given energy), with the following boundary conditions
\begin{equation}\label{eq:bound}
\psi_\text{sc}(\vb*{r})=\begin{cases}\phi_n^{\text{R}*}(\vb*{r})+\sum_{m\neq n}B_m^\text{R}\phi_m^\text{R}(\vb*{r}), & x\to\infty\\\sum_{m}A_m^\text{L}\phi_m^{\text{L}*}(\vb*{r}), & x\to-\infty\end{cases}
\end{equation}
where $\phi_n^{\text{L/R}}(\vb*{r})$ and $\phi_n^{\text{L/R}*}(\vb*{r})$ are the left- and right-going running wave states in the left/right electrodes with energy $E$. The transmission and reflection coefficient can be calculated by
\begin{eqnarray}
T_n(E)=\frac{\sum_m|A_m^\text{L}|^2[\text{d}E_m^\text{L}(k)/\text{d}k]|_{k_m^\text{L}}}{[\text{d}E_n^\text{R}(k)/\text{d}k]|_{k_n^\text{R}}}\\
R_n(E)=\frac{\sum_{m\neq n}|B_m^\text{R}|^2[\text{d}E_m^\text{R}(k)/\text{d}k]|_{k_m^\text{R}}}{[\text{d}E_n^\text{R}(k)/\text{d}k]|_{k_n^\text{R}}}
\end{eqnarray}
To solve the scattering states, we introduce some ``system states'' $\psi_l$ which satisfy the equation
\begin{equation}\label{eq:wl}
(\hat{H}-E)\psi_{l}(\vb*{r})=w_{l}(\vb*{r})
\end{equation}
where $w_l$ is a perturbation function that is nonzero only near the boundary far from the central region. The $\psi_l$ are decomposed to left and right electrode running wave states $\phi_m^{\text{L}}$ and $\phi_m^{\text{R}}$ at one left electrode unit cell $\Omega_\text{L}$ and one right electrode unit cell $\Omega_\text{R}$. Using such decomposition, we can then use linear combination of $\psi_l$ to generate the scattering states which satisfy the boundary conditions Eq. (\ref{eq:bound}) (where the certain running wave coefficients are zero). We can then get $T_n(E)$ and thus solve the transport problem. More details of the original method can be found in Ref. \cite{GARCIALEKUE2015292}.

\subsection{Electronic structure}
The above approach can be easily implemented in the common DFT codes, and its computational time is similar to that of a standard ground-state calculation. A package called PEtot\_trans\cite{PWtrans} has been developed using this approach based on PEtot code\cite{PEtot}. But common DFT algorithm has the computational complexity of $O(N^3)$, that makes it very expensive for large systems. In practice, we can only calculate the electronic structure of a system with hundreds of atoms using the above approach. Fortunately, several linear-scale DFT algorithms have been developed.\cite{PhysRevB.52.1640,PhysRevLett.73.122,PhysRevB.47.10891,PhysRevB.47.9973,PhysRevB.48.14646,PhysRevLett.66.1438} Those algorithms have the computational complexity of $O(N)$, enable them for large system calculations. One such algorithm is called linear-scaling three-dimensional fragment (LS3DF) method, and has been used to study the electronic structure of the systems containing more than 10000 atoms based on plane wave expansions.\cite{Wang2008,0953-8984-20-29-294203,Wang2014} This is benefited from the highly-scalable divide-and-conquer scheme implemented in the LS3DF code, which makes it not only linear scaling to the system size, but also linear scaling to the number of processors used in the computations.
\begin{figure}[htbp]
\centering
\includegraphics{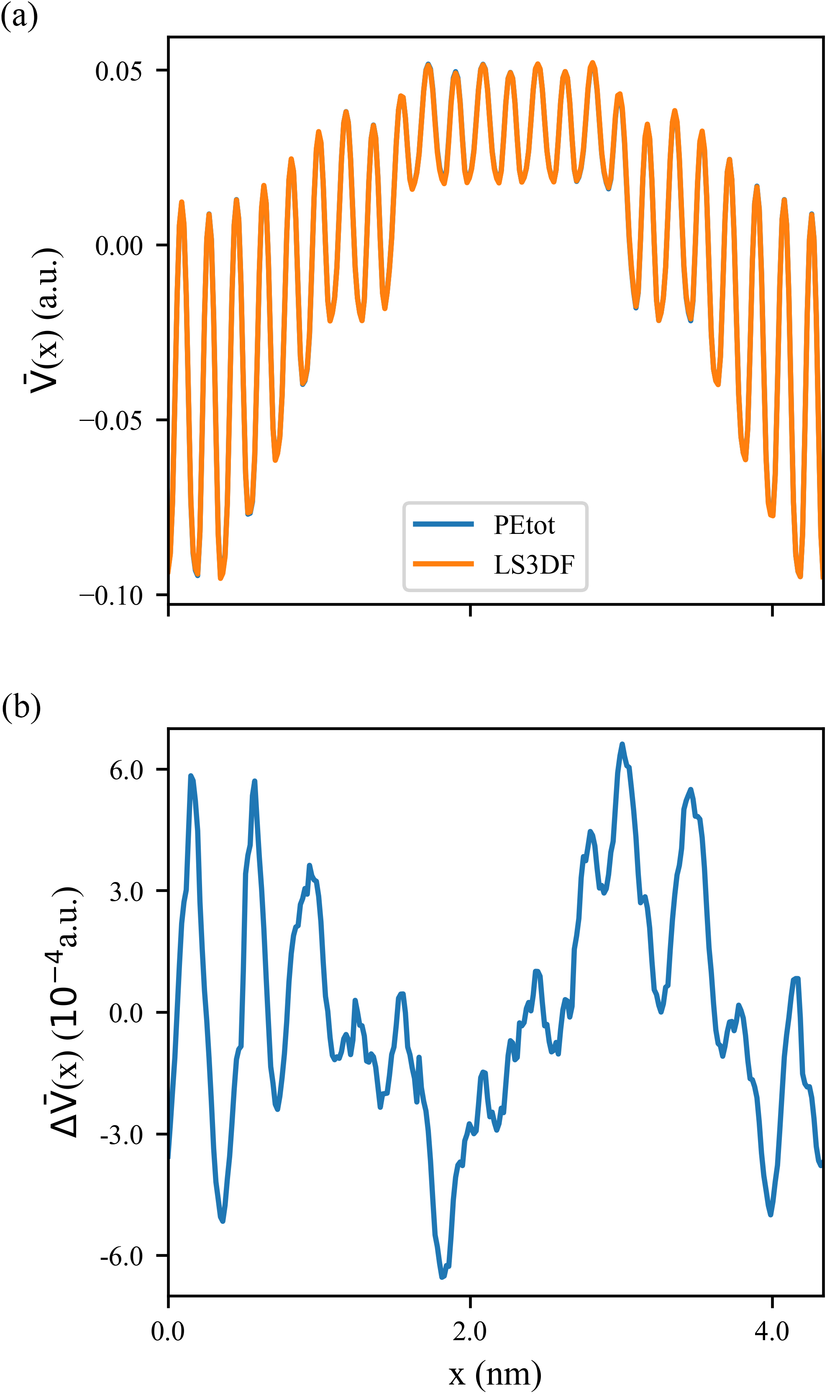}
\caption{(a) is the average potential on the $yz$-plane calculated by LS3DF (the orange line) and PEtot (the blue line). (b) is the error of average potential between the calculations of these two codes.}\label{fig:vr-compare}
\end{figure} 

In this work, we use a GPU version of LS3DF code\cite{JIA20178}. Like most linear scaling DFT method, LS3DF is based on the nearsightedness of the DFT properties. Such nearsightedness is rigorous for semiconductors,\cite{PhysRevLett.76.3168} but it is not clear whether LS3DF will still work for metallic system. We thus need first to check whether LS3DF works for the metallic system. Here we use a relatively small system of 488 copper atoms, whose structure is similar to those we study, to test the validity of LS3DF and its speed. We find that it takes about 500 minutes if we use the common Kohn-Sham DFT code (PEtot) with 360 CPU processors (5760 cores), but only 30 minutes if the GPU version of LS3DF code and 480 GPUs are used. In the LS3DF calculation, the system has been divided into  95 overlapping fragments, and each fragment contains $9\sim334$ copper atoms without any surface passivation. To compare the results calculated by these two software, we calculate the average potential on the $yz$-plane ($\bar{V}(x)$) and show them in FIG. \ref{fig:vr-compare}(a). We can find that the error between these two method are less than $\SI{0.02}{eV}$ (as shown in FIG. \ref{fig:vr-compare}(b)), which indicated that LS3DF gives almost the same result as PEtot and thus suggested that LS3DF works well for the metallic system. 

\begin{figure}[htbp]
\centering
\includegraphics{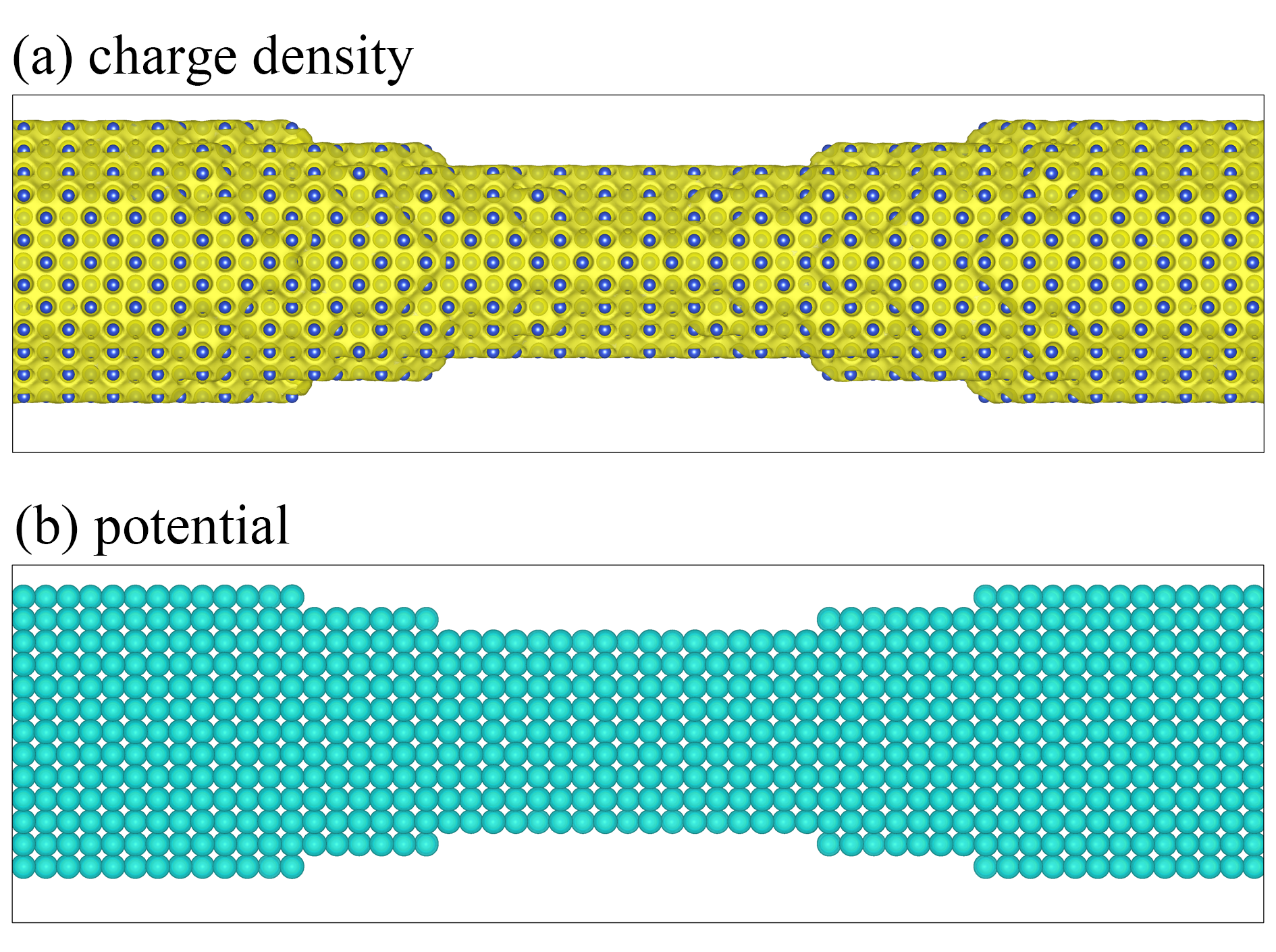}
\caption{The (a) charge density and (b) potential of the central region of system III calculated by LS3DF code. The isovalues are 0.01 and 0.5 a.u. respectively.}\label{fig:ls3df}
\end{figure}

In LS3DF calculation, a large system is first divided into small overlapping fragments. The wave functions and charge densities of individual fragments are calculated separately by specific groups of computer processor. These fragment charge densities are then patched together to yield the global charge density using a special patching scheme, The Poisson equation of the whole system is solved based on the patched global charge density. For a metallic system, a global Fermi energy is used to control the orbital occupations of all fragments. For a given accuracy, the size of the fragment will be fixed regardless of the total system size. The larger system just means larger number of fragments. Besides, larger number of computer processor means small number of fragments each computer processor group has to deal with. This leads to a double linear scaling property of LS3DF: linear both to the size of the system, and number of computer processor used. Due to the double linear scaling, LS3DF can calculate the large systems quickly. For instance, it only takes about 30 minutes to calculate the charge density of the central region of the system III (which contains about 3000 atoms) using 1920 GPUs, and the results are shown in FIG. \ref{fig:ls3df}.

\subsection{The folded spectrum method and acceleration technique}
After we solved the problem of LS3DF for  self-consistent field solution of the charge density and potential of the system, we now turn our attention to solve the scattering state of the system. The preconditioned conjugate gradient (PCG) method can be quite efficient to solve the linear equation Eq. (\ref{eq:wl}). Unfortunately, $(\hat{H}-E)$ is not positive definite, so the PCG method cannot be directly applied. In the original approach, we have calculated all the eigenstates of $\hat{H}$ below the given energy $E$ and depleted those eigenstates from $\hat{H}$ to make $(\hat{H}-E)$ positive definite. However, this approach only works for small system where all eigenstates below $E$ can be solved using iteration methods. For a system contains thousands of atoms, there could be tens of thousands of eigenstates below $E$ (which is usually near the Fermi level of electrodes), so it will be very expensive to calculate all of them. This becomes the main limitation to solve the scattering states using plane wave basis set.

To solve this problem, we use the folded spectrum method\cite{doi:10.1063/1.466486}, which apply another $(\hat{H}-E)$ on both sides of Eq. (\ref{eq:wl}), then it becomes
\begin{equation}\label{eq:wl1}
(\hat{H}-E)^2\psi_{l}(\vb*{r})=(\hat{H}-E)w_{l}(\vb*{r})=w'_l(\vb*{r})
\end{equation}
Since $(\hat{H}-E)^2$ is always positive definite, this equation can be directly solved by using the PCG method. For the PCG method, we use PCG to minimize the following quantity:
\begin{equation}
F=\frac12\expval{(\hat{H}-E)^2}{\psi_l}-\braket{\psi_l}{w_l}
\end{equation}
For some system, such as a 2000 atoms black phosphorus nanosheet, the PCG iteration converges in several thousands steps when solving Eq. (\ref{eq:wl1}), thus it can be directly used. But in many cases, such as the copper system we study here, the PCG iteration is very hard to converge. This is due to the poor condition number of operator $(\hat{H}-E)^2$, especially when there are eigenstates with eigenenergy $\varepsilon_i$ close to $E$. For the four copper systems we study, more than 30000 steps of PCG is required to converge (which means the error, $\norm{(\hat{H}-E)\psi_l-w_l}$, is less than \num{1e-8}, in atomic unit). This is rather expensive. Especially, since there is no way to massively parallelize the PCG iterations, the calculation takes a long time, without the benefits of massive supercomputer.

The solution of Eq. (\ref{eq:wl1}) can be expanded by all the eigenstates of $\hat{H}$
\begin{equation}
\psi_l=\sum_i\frac{\braket{\phi_i}{w_l'}}{(\varepsilon_i-E)^2}\phi_i
\end{equation}
Here $\phi_i$ and $\varepsilon_i$ are the $i$-th eigenstate and eigenvalue of $\hat{H}$. Thus the eigenstates around the given energy $E$ contribute the most to $\psi_l$. If we have these eigenstates, they can be used to accelerate the PCG convergency (for instance, they can be depleted from the Hamiltonian). In this case, those eigenstates are just used to accelerate the convergency, there is no need to calculate all the eigenstates below $E$, thus we need much smaller number of eigenstates.

\begin{figure}[htbp]
\centering
\includegraphics{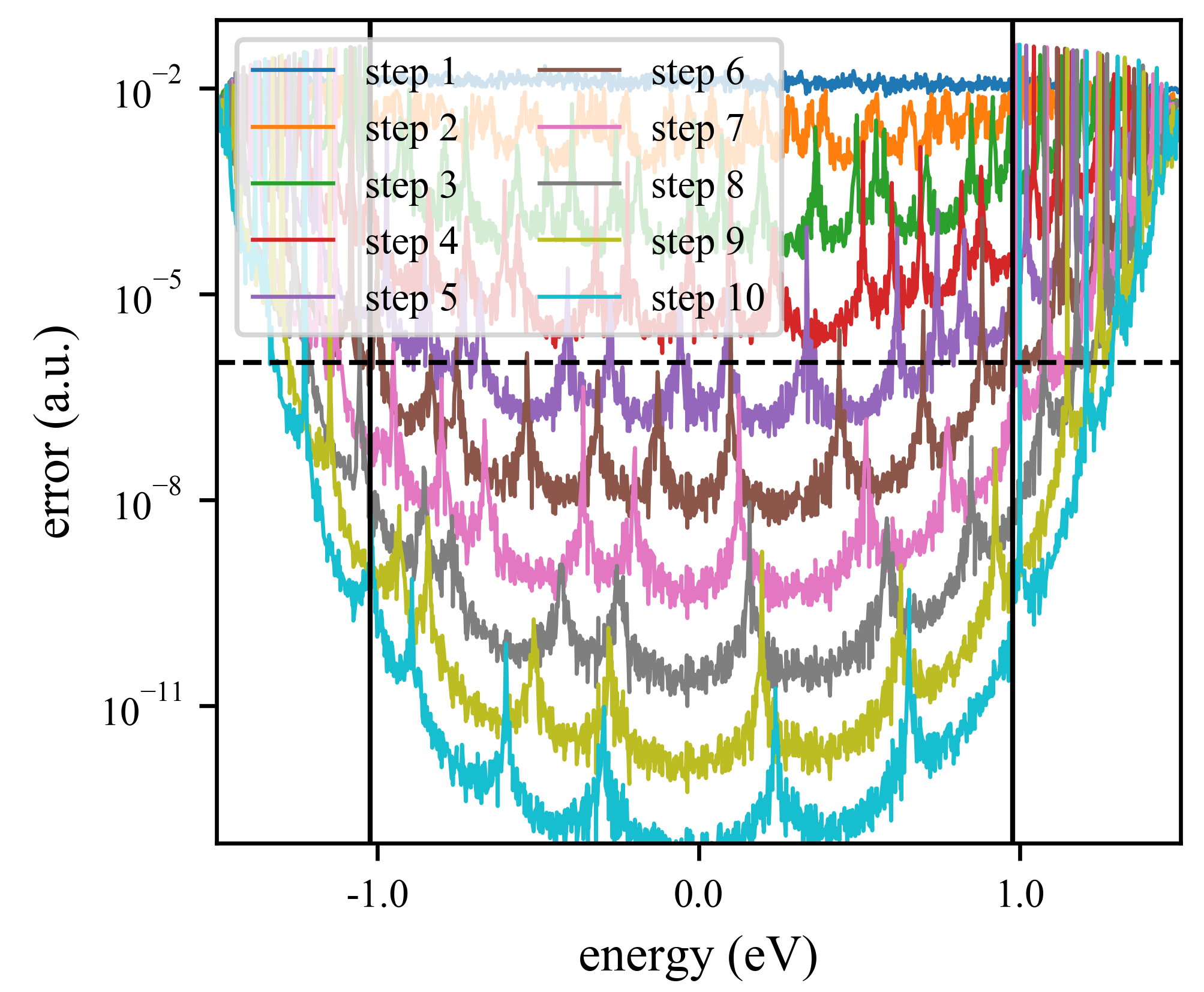}
\caption{Error of each eigenstates when a 1500-degree ChebFD is applied to system III, step by step. Here we ignore some states with very large error, which are called ``ghost states''. The boundary of the energy window and the tolerence of error we use in most calculations are marked by two black solid lines and one black dash line, respectively.}\label{fig:chebfd}
\end{figure}

To get a limited number of eigenstates around $E$, there are several methods one can use, one is the folded spectrum method\cite{doi:10.1063/1.466486}. However, we meet the same converge problem as solving Eq. (\ref{eq:wl1}). Here we adopted the Chebyshev filter diagonalization (ChebFD) algorithm\cite{Weise2006,PIEPER2016226}, to find the eigenstates within a energy window around $E$. In the ChebFD algorithm, a polynomial of the Hamiltonian $\hat{H}$, which is constructed by the linear combination of Chebyshev polynomials, is used to imitate a window function surrounding $E$, and this polynomial $p(\hat{H})$ can be viewed as a filter. Then this filter can be applied to a group of random wave functions $\{\phi_i^{(0)}\}$. Assume $\phi_i^{(0)}=\sum_jc_{ij}\phi_j$, where $\phi_j$ is the eigenstate of $\hat{H}$ with eigenenergy $\varepsilon_j$, then $\phi_i'^{(0)}=p(\hat{H})\phi_i^{(0)}=\sum_jc_{ij}p(\varepsilon_j)\phi_j$, thus it will suppress the amplitudes of eigenstates with their $\varepsilon_j$ outside the window, while keeping the eigenstates inside the window. One can then use $\{\phi_i'^{(0)}\}$ to do a subspace diagonalization of $\matrixel{\phi_i'^{(0)}}{\hat{H}}{\phi_j'^{(0)}}$, yield eigenstates $\{\phi_i^{(1)}\}$ of this subspace. This is one ChebFD step, and one can repeat this ChebFD step by applying $p(\hat{H})$ again to $\{\phi_i^{(1)}\}$. Several ChebFD iterations need to be used to get accurate eigenstates in practice. One major benefit of ChebFD algorithm is its parallelization. Because $p(\hat{H})\phi_i^{(n)}$ can be done independently for different $\phi_i^{(n)}$, the procedure is embarrassingly parallel. Although this algorithm may loss some eigenstates in the energy window, this leakage is not important for our case, since we only use them as preconditioner. FIG. \ref{fig:chebfd} shows how the ChebFD algorithm works for system III, the largest one in our test systems. Here a 1500-degree of ChebFD filter (1500 order in $p(\hat{H})$ polynomial) and 1500 random initial wave functions are used. The energy window is set to \SI{2.0}{eV}. As shown in FIG. \ref{fig:chebfd}, the overall error of the eigenstates decrease by about one order of magnitude with each step of ChebFD. Here we use 43200 CPU cores on Titan, and it takes about half an hour to run one step of ChebFD. In doing so, we have used 432 CPU cores for each $p(\hat{H})$. and the total number of processors have been divide into 100 groups. In this part of calculations, only CPU codes are available. We expect future implementation on GPU nodes can significantly accelerate this calculation, and make such calculation more routine.

When we get the eigenstates around $E$, we can use them to construct a projector and than apply it to Eq. (\ref{eq:wl1}) (deplete out such eigenstates), as we did in the original approach. But that require all the eigenstates are fully converged. Here, instead, we use them as preconditioner. As we know, a good preconditioner can significantly improve the convergency of PCG iterations. We have already used one $G$-space diagonal preconditioner. but it is only effective on the high energy plane wave at the higher end of $(\hat{H}-E)^2$. Now we have some eigenstates around $E$, we can use them to construct a much better preconditioner (which is closer to $(\hat{H}-E)^{-2}$)
\begin{equation}
\begin{split}
M'&=M+\sum_{i=1}^{N}\left[\frac{1}{(\varepsilon_i-E)^2+\Delta_i^2}-M\right]\ketbra{\phi_i}\\
&=M\left(1-\sum_{i=1}^N\ketbra{\phi_i}\right)+\sum_{i=1}^{N}\frac{\ketbra{\phi_i}}{(\varepsilon_i-E)^2+\Delta_i^2}
\end{split}
\end{equation}
here $M(G)=\frac{1}{(G^2-E)^2+\Delta E^2}$ is our $G$-space diagonal preconditioner, and $M'$ is our new preconditioner. Basically, in the subspace $\spn\{\ketbra{\phi_1},\ketbra{\phi_2},\cdots,\ketbra{\phi_N}\}$, we use the $(\hat{H}-E)^{-2}$ inversion, while in the rest of the space, we use the $M$ preconditioner. This formalism looks almost the same as the projector one. But in the current approach, the eigenstates $\{\phi_i\}$ need not to be very accurate. In particular, we have $\Delta_i=\norm{\hat{H}\phi_i-\varepsilon_i\phi_i}$ is the error of the eigenstate $\phi_i$. In this way, we can use the approximated eigenstates to accelerate the PCG convergency.

\begin{figure}
\centering
\includegraphics{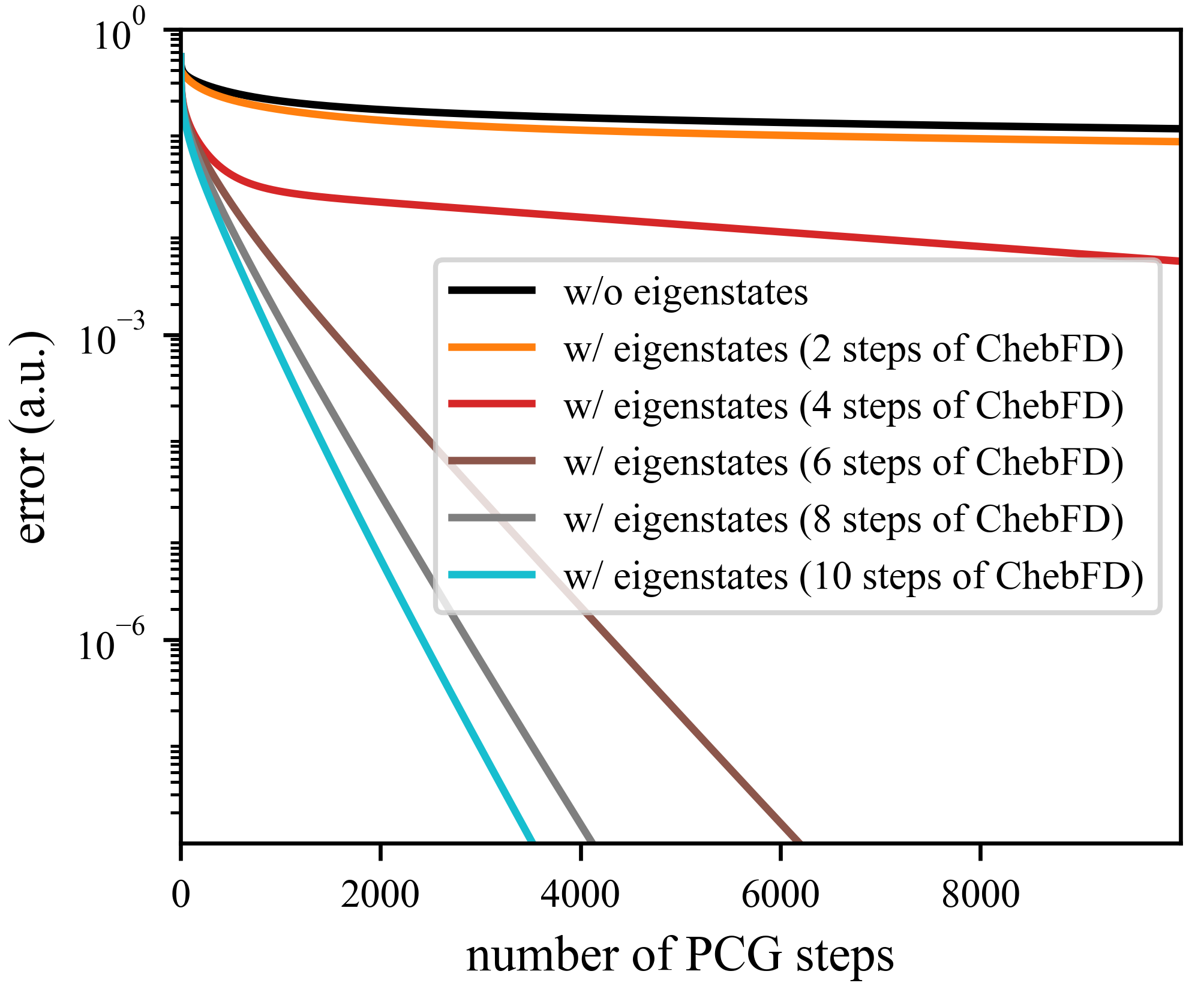}
\caption{The convergence speed of PCG progress for system III. The black line represents the PCG progress without eigenstates, while the orange, red, brown, gray, and cyan line represent the PCG using eigenstates generated by 2, 4, 6, 8, and 10 steps of ChebFD (as shown in FIG. \ref{fig:chebfd}), respectively.}\label{fig:converge}
\end{figure} 

Here we also use system III to test our accelerate technique, all the results are shown in FIG. \ref{fig:converge}. For that system, the linear equation error defined as $\norm{(\hat{H}-E)\psi_l-w_l}$ decreases very slowly if no eigenstates are used for acceleration, and it is still larger than \num{0.1} (in atomic unit) after 10000 steps of PCG. Then we use eigenstates with different accuracy (which means generated by different steps of ChebFD, as shown in FIG. \ref{fig:chebfd}) to construct the preconditioner and then repeat the PCG iterations. Unfortunately, we find that eigenstates generated by 2 or 4 steps of ChebFD only accelerate the PCG convergency modestly, which suggests that eigenstates with low accuracy ($\Delta_i>\num{1e-5}$) cannot  be used for effective acceleration. Only when the eigenstates with $\Delta_i\le\num{1e-6}$ are used in the calculations, and we find the iteration has been accelerated significantly. It takes about 6200, 4100, and 3500 PCG steps to suppress the linear equation error below \num{1e-8}, when using eigenstates generated by 6, 8, and 10 steps of ChebFD, respectively. As shown in FIG. \ref{fig:chebfd}	, $\Delta_i$ of most eigenstates used here are less than \num{1e-6}, \num{1e-8}, and \num{1e-10} in these three cases. It takes only 2-3 seconds for one PCG step using 432 CPU cores, so the solving of the linear equation itself is not expensive if we already have the eigenstates. Our result also shows that the error of eigenstates (\num{1e-6} for this system) can be greater than the linear equation error (\num{1e-8} for this system), which will not be true in the projector depletion approach. This shows the advantage of using preconditioner approach.

\subsection{Wave function interpolation technique}
As discussed above, by using the LS3DF and folded spectrum method (with acceleration), we have solved two main problems to apply plane-wave transport method to large system.  Nevertheless, there are still some detailed issues in practice worth to be discussed.

In a large system, the electronic structure of the electrodes can be quite complex. Since we need to calculate the running-wave states in the electrode, the band structure with many hundreds of $k$-points along the transport direction (which is $x$-direction in this work) is required. It turns out to be very expensive to do such calculations with so many $k$-points for an electrode with hundreds of atoms. Here we employ wave function interpolation to extract the eigenstates between two nearest $k$-points which have already been calculated. In this way, we can use relatively small number of $k$-points to get the accurate whole band structure. The details of the wave function interpolation technique are discussed in the following.

If we have already calculated the orthonormal eigenstates of two nearest $k$-points, $k_1$ and $k_2$, which means
\begin{eqnarray}
\hat{H}{(k_{1,2})}\ket{\psi_i{(k_{1,2})}}=\varepsilon_i{(k_{1,2})}\ket{\psi_i{(k_{1,2})}}
\end{eqnarray}
Since  $k_1$ and $k_2$ are very close, so for a $k$ in $(k_1,k_2)$, the Hamiltonian $\hat{H}(k)$ can be expressed using a linear interpolation
\begin{equation}
\hat{H}(k)\approx\alpha\hat{H}(k_1)+(1-\alpha)\hat{H}(k_2)
\end{equation}
where $\alpha=\frac{k-k_2}{k_1-k_2}$. So the matrix elements of $\hat{H}(k)$ can be written as
\begin{equation}\label{eq:Hk}
\begin{aligned}
&{H}_{ij}(k)=\matrixel{\psi_i(k_1)}{\hat{H}(k)}{\psi_j(k_1)}\\
&=\alpha\varepsilon_i(k_1)\delta_{ij}+(1-\alpha)\matrixel{\psi_i(k_1)}{\hat{H}(k_2)}{\psi_j(k_1)}
\end{aligned}
\end{equation}
Here a transformation matrix $U$ between $\psi_i(k_1)$ and $\psi_j(k_2)$ can be used, which means $\psi_i{(k_1)}=\sum_jU_{ij}\psi_j(k_2)$. If all the eigenstates are included, $U$ will be the overlap matrix $U_{ij}=S_{ij}=\braket{\psi_j(k_2)}{\psi_i(k_1)}$. In reality, only finite number of $\psi_i$ are used, thus the basis is incomplete, and $S$ is not a unitary matrix. A Gram-Schmidt process on $S$ is used to make a unitary $U$ out of $S$ matrix. Under this approximation, we can rewrite Eq. (\ref{eq:Hk}) as
\begin{equation}
H_{ij}(k)=\alpha\varepsilon_i(k_1)\delta_{ij}+(1-\alpha)\sum_{m}U_{im}^*U_{jm}\varepsilon_m(k_2)
\end{equation}
Then we can get the eigenstates $\psi_i(k)$ and $\varepsilon_i(k)$ by diagonalizing of $H(k)$, and $\psi_i(k)=\sum_jc_j\psi_j(k_1)$. However, this formula will have a jump at $k_2$. To make it continuous, we can expand $\psi_i(k)$ by both $\psi_i(k_1)$ and $\psi_i(k_2)$, such as
\begin{equation}
\psi_i(k)=\alpha\sum_jc_j\psi_j(k_1)+(1-\alpha)\sum_{j,k}c_jU_{jk}\psi_k(k_2)
\end{equation}

\begin{figure}[htbp]
\centering
\includegraphics{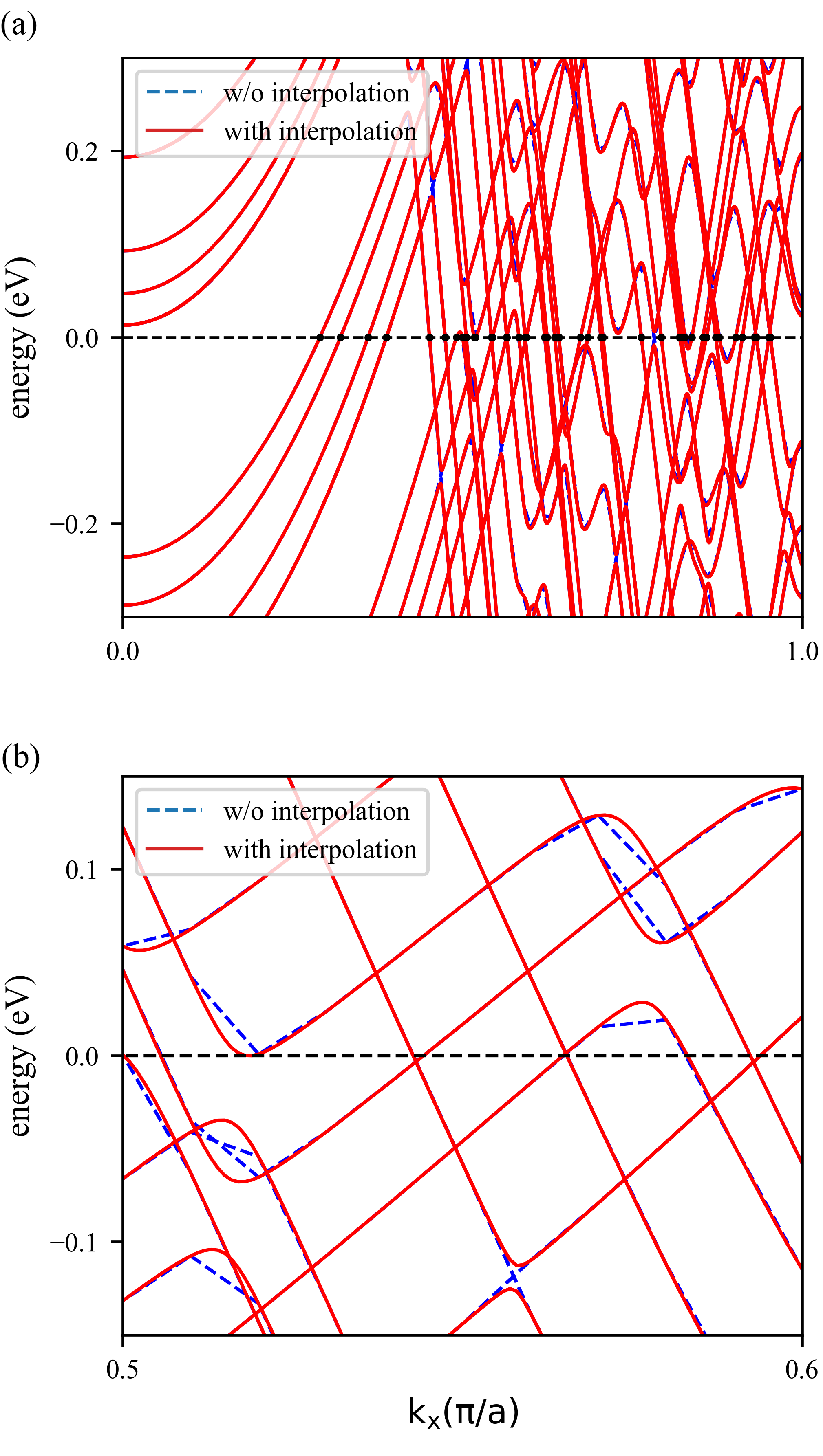}
\caption{The (a) full and (b) detail band structure without (blue dash line) and with (red solid line) wave function interpolation along the $x$-direction. The Fermi level has been adjusted to zero-point of energy in both figures. The black dots in (a) shows the $k(x)$ of the running waves at the Fermi level.}\label{fig:bands}
\end{figure}

As we discuss above, all the system we study have the same structure of electrode, which contains 145 copper atoms per unit cell. In the ground-state calculation, we use 101 $k$-points along the $x$-direction and output all the eigenstates in real space. Then we use the above wave function interpolation technique, to add 9 $k$-points between every two nearest $k$-points. In this way, we get the band structure of 1001 $k$-points along the $x$-direction.

The band structure without and with wave function interpolation are shown in FIG. \ref{fig:bands}, which are represented by the blue dash and red solid line, respectively. It is easy to find that those two band structures are the same in most parts, but have some differences when it is blewd up. Overall, the bands with interpolation are much smoother than those without interpolation. We also find that there are some crossing bands in the original band structure, which become anitcrossing bands after interpolation. In the following, we will get running-wave states with higher accuracy using bands with interpolation.

Since the system we study are all metallic structures, the transport properties near the Fermi level of electrode are the most important. At the Fermi level (which has been adjusted to zero in FIG. \ref{fig:bands}), there are 52 running-wave states (their conjugate states are not counting), but this number changes to 50 after interpolation, as shown in FIG. \ref{fig:bands}(a). This is due to the detailed difference between those two band structures, especially the anticrossing bands, which is shown in FIG. \ref{fig:bands}(b). Each of these running wave states $\phi_i$ with energy of $E$ will be used to construct one $w_i$ by multiplying $\phi_i(\vb*{r})$ with a mask function $m(x)$, which is nonzero only near the artificial boundary of the system, far away from the central part of the interconnect.

\section{RESULTS AND DISCUSSION}
\begin{figure*}[htbp]
\centering
\includegraphics{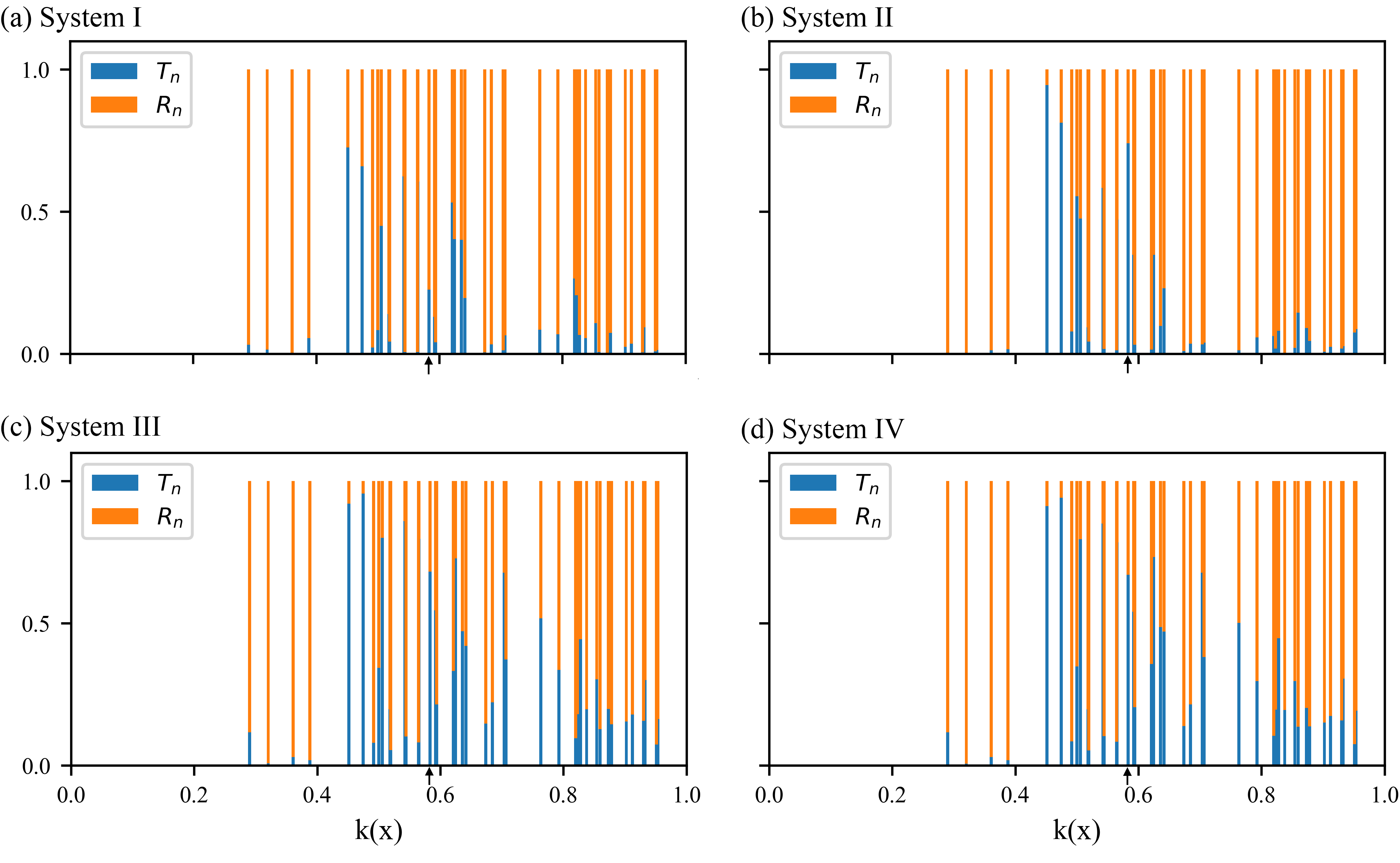}
\caption{The (normalized) transmission and reflection coefficient of different incoming running wave states, in (a)-(d) system I-IV. The blue and orange bar represent the transmission and reflection coefficient, respectively, while the vertical uparrow indicate the scattering states we study.}\label{fig:TR}
\end{figure*}

Finally, we could use the calculated system states $\psi_l$ to construct scattering state $\psi_\text{SC}$ and thus solve the transport problem. Since all the three systems are metal, only their transport properties at the Fermi level of the electrodes are studied in this work. The transmission ($T_n$) and reflection coefficient ($R_n$) of each scattering states $\psi_\text{sc}(n)$ are shown in FIG. \ref{fig:TR}. Note we have made $T_n+R_n=1$.

As we discussed above, there are 50 left-going running wave states from the right electrode at the Fermi level (the black dots in FIG. \ref{fig:bands}(a)). FIG. \ref{fig:TR} shows the transmission and reflection coefficient of the those incoming electrode states in the systems we studied. We find that there are always some states with relatively large $T_n$ (for example, the states with $k(x)=0.451$ and 0.475), while the magnitudes of $T_n$ of some states are greatly influenced by the ``neck'' structure of the system. Overall, system I and and II have similar transmission patterns than system III and IV, indicating the importance of the narrowest neck in controlling the transmission.

\begin{figure*}[htbp]
\centering
\includegraphics{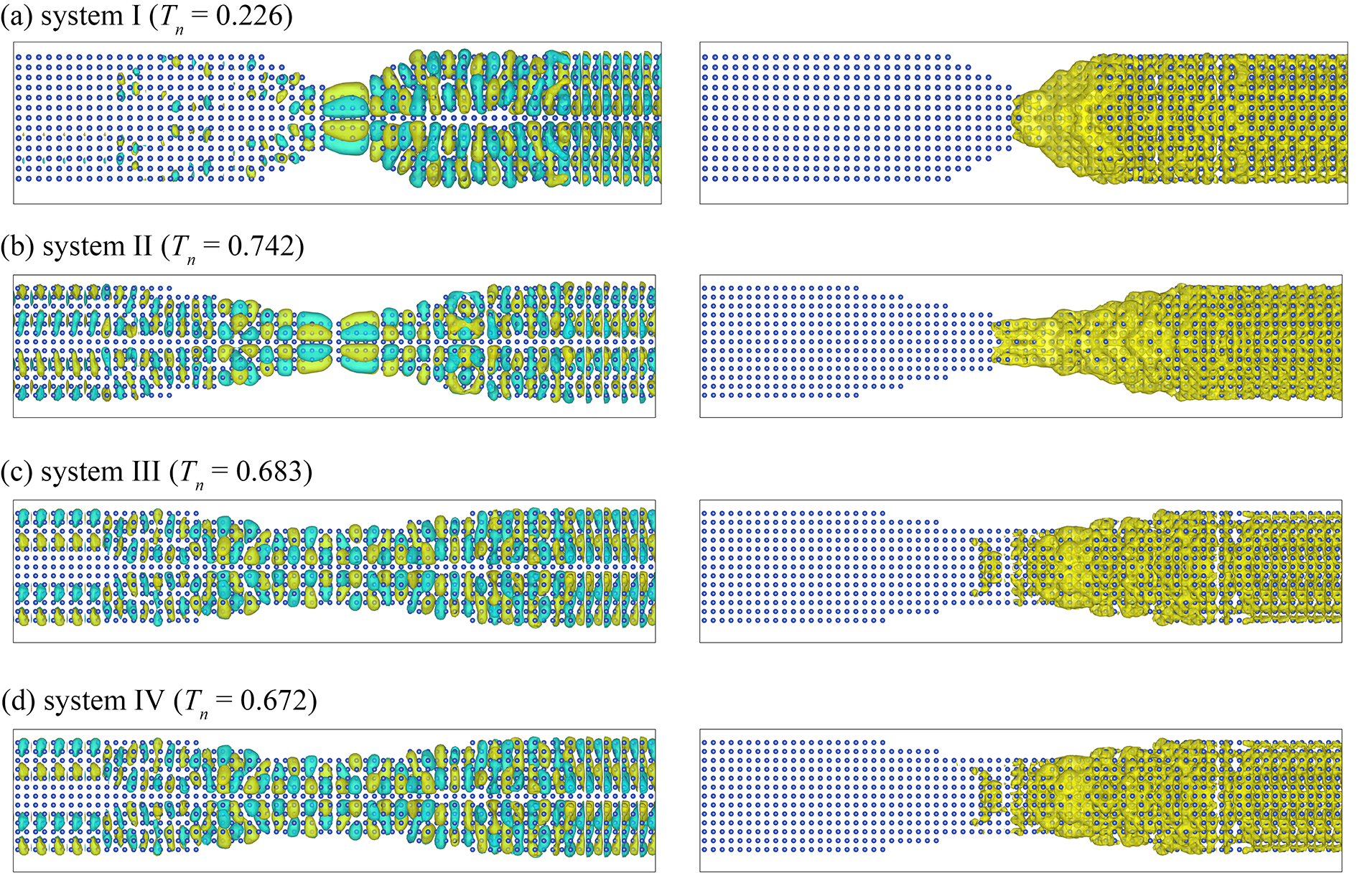}
\caption{The (left panel) scattering state wave function with the same right electrode incoming state and (right panel) charge density at the Fermi level of the electrode, in (a)-(d) system I-IV . The isosurface is set to 0.01 and 0.008 a.u., for wave function and charge density, respectively.}\label{fig:wf}
\end{figure*}

The left panel in FIG. \ref{fig:wf} shows the scattering wave functions of the four interconnect from the same right-incoming wave (with $k(x)=0.583$) as indicated by the vertical arrow shown in FIG. \ref{fig:TR}. We find that the detailed structure in the central region (where the scattering occurs) influences the scattering wave functions significantly. System II, III, and IV have more components of wave function in the left electrode than that of system I, therefore their $T_n$ is much larger. We have also calculated the charge density  of all the scatter states wave functions at the Fermi energy
\begin{equation}
\rho_E(\vb*{r})=\sum_{i=1}^N|\psi_\text{sc}(\vb*{r})|^2
\end{equation}
Here $N$ is the number of left-going running-wave states in the right electrode. These are shown in the right panel of FIG. \ref{fig:wf}. Since $\rho_E(\vb*{r})$ mainly distribute in the right side of all the systems we study, it suggests that the overall transmission coefficient of those system is not very large.

However, it is hard to find how the shape of central part influences $T_n$ for each individual scattering states, as we discussed above. But in general, system III and IV have larger $T_n$ than I and II. We can define the average transmission coefficient $T=\sum_{i=1}^NT_i/N$ to show this tendency. This $T$ also represents the overall conductivity of this interconnect. As shown in TABLE \ref{tab}, the average transmission coefficient of system I and II are very close, which is about half of that of system III and IV. This $T$ increases with the ``neck'' diameter $r$. However, the $T/r^2$ ratio is larger for smaller neck interconnect, indicating the quantum mechanical effect, might enhance the transmission on top of the simple classical estimations. The other geometry parameter $l$, which measure the length of the ``neck'' region, is not so sensitive compare to the radius of the ``neck'' $r$ in terms of controlling $T$. $T$ only decreases slightly when $l$ increases from \num{2.0} to \SI{5.0}{nm}.

\begin{table}[htbp]
\caption{Number of atoms ($N_\text{Cu}$), the radius ($r$) and length ($l$) of the ``neck'' region, and the average transmission coefficients at the Fermi level of the electrodes ($T$) of the system I-IV.}\label{tab}
\begin{ruledtabular}
\begin{tabular}{*6c}
 system & $N_\text{Cu}$ & $r$ (\si{nm}) & $l$ (\si{nm}) & $T$ & $T/r^2$ (\si{nm^{-2}})\\\hline
 I & 3996 & 0.5 & 2.0 & 0.179 & 0.716\\
 II & 3660 & 0.5 & 5.0 & 0.182 & 0.728\\
 III & 4140 & 0.75 & 5.0 & 0.350 & 0.622\\
 IV & 4139 & 0.75 & 5.0 & 0.352 & 0.626\\
\end{tabular}
\end{ruledtabular}
\end{table}

In system IV, we introduce a vacancy defect in system III, as shown in FIG. \ref{fig:wf}(d). We find that it only influences the scattering state of the same channel very slightly. Actually, it increases the overall $T$ slightly. It is rather surprising for such small influence given the small size of the neck. This phenomenon suggests that some point defect might not have detrimental effects on the transmission. All these phenomenon warrant more comprehensive future investigations. 

\section{CONCLUSION}
In summary, we have improved a previously developed \textit{ab initio} quantum transport simulation method based on plane wave basis set. By introducing the linear-scale ground-state DFT algorithm (LS3DF), folded spectrum method (with acceleration), and other techniques such as filtering and wave function interpolation, we are capable to simulate the transport properties of a system with several thousands of atoms. Since the computational complexity of our new method is nearly linear to the scale of system, the system scale will be able to be extended to tens of thousands with massively parallelize. The future implement on GPU nodes is also expected to accelerate the calculation. 

By using this new method, we study the transport properties of several copper nanowire systems with about 4000 atoms, and show the shape and point defect effects in them. We find that the conductance of copper nanowire is mainly determined by the radius of narrow part, not the length, and some point vacancy defect might even improve the transmission. These simulations suggest the potential usage of our method for the study and design of nanoscale interconnect in the advanced electronic devices.

\textit{Ab initio} simulation is useful to investigate and design the nano-electronic devices. Our method provide a new approach to study  the transmission phenomenon in them. However, real semiconductor device, such as transistor, is more complex than then interconnect system we study here. Further development is expected to make our method can be applied to real devices with tens of thousands of atoms.

\begin{acknowledgments}
L.W.W. was supported by the US Department of Energy, Office of Science, Office of Basic Energy Sciences, Materials Sciences and Engineering Division under contract no. DE-AC02-05-CH11231 within the beyond-Moore's law LDRD project. M.Y., X.J., and S.S.L. were supported by China Key Research and Development Program (2018YFA0306101), National Natural Science Foundation of China (Grand Nos. 11774338, 11574304), Chinese Academy of Sciences-Peking University Pioneer Cooperation Team (CAS-PKU Pioneer Cooperation Team), the Youth Innovation Promotion Association CAS (Grand No. 2016109).  

This work used the resources of the National Energy Research Scientific Computing Center (NERSC)  and Oak Ridge Leadership Computing Facility (OLCF) through the Innovative and Novel Computational Impact on Theory and Experiment (INCITE) project NTI009.
\end{acknowledgments}

\bibliography{transport}

\end{document}